\documentclass[%
 aip,
 jmp,%
 amsmath,amssymb,
preprint,nofootinbib%
]{revtex4-1}

\usepackage{graphicx}
\usepackage{dcolumn}
\usepackage{bm}
\usepackage{amsfonts}
\usepackage{color}

\usepackage{ccaption,subfig}

\usepackage[export]{adjustbox}
\usepackage{floatrow}

\usepackage{float}
\floatstyle{plaintop}
\restylefloat{table}

\usepackage{footmisc}
{
\makeatletter
\def\frontmatter@thefootnote{%
 \altaffilletter@sw{\@fnsymbol}{\@fnsymbol}{\csname c@\@mpfn\endcsname}%
}}%

\makeatother

\begin{document}

\title{On the role of mechanical feedback in synchronous to asynchronous transition during embryogenesis}
\author{Abdul N. Malmi-Kakkada$^{*}$}
\email{amalmikakkada@augusta.edu}
\affiliation{Department of Physics and Biophysics, Augusta University, Augusta, GA 30912, USA}

\author{Sumit Sinha$^{*,\#}$}
\email{ssinha@g.harvard.edu}
\affiliation{Department of Physics, University of Texas, Austin, TX 78712, USA}

\author{D. Thirumalai}
\email{dave.thirumalai@gmail.com}
\affiliation{Department of Physics, University of Texas, Austin, TX 78712, USA}
\affiliation{Department of Chemistry, University of Texas, Austin, TX 78712, USA}

\def\thefootnote{*}\footnotetext{Equal Contribution}\def\thefootnote{\arabic{footnote}}

\def\thefootnote{\#}\footnotetext{Present Address: Applied Mathematics (SEAS, Harvard University) \& Data Science (Dana Farber Cancer Institute)}\def\thefootnote{\arabic{footnote}}


\date{\today}

\begin{abstract}
Experiments have shown that during the initial stage of Zebrafish morphogenesis a synchronous to asynchronous transition (SAT) occurs,  as the  cells  divide extremely rapidly. In the synchronous phase, the cells divide in unison unlike in the asynchronous phase. Despite the wide spread observation of SAT in experiments, a theory to calculate the critical number of cell cycles, $n^{*}$, at which asynchronous growth emerges does not exist. Here, using a model for the cell cycle, with the assumption that cell division times are Gaussian distributed with broadening, 
we predict $n^{*}$ and the time at which the SAT occurs. The theoretical results are in excellent agreement with experiments. The theory, supplemented by agent based simulations, establish  that the SAT emerges as a consequence of biomechanical feedback on cell division. The emergence of asynchronous phase is due to linearly increasing fluctuations in the cell cycle times with each round of cell division. We also make several testable predictions, which would further shed light on the role of biomechanical feedback on the growth of multicellular systems. 
\end{abstract}


\maketitle


Time dependent synchronous behavior is pervasive across biological 
systems. Numerous examples of synchronous behaviors exist, from networks of 
neurons in the circadian pacemaker~\cite{enright1980temporal,mirollo1990synchronization}, and insulin-secreting pancreatic 
cells~\cite{sherman1988emergence} at the cellular scale, to the chirping of crickets~\cite{walker1969acoustic}, and synchronous flashing of a colony of 
fireflies~\cite{smith1935synchronous,sarfati2021self}. A particularly fascinating example of tissue scale synchronous behavior occurs during 
the development of multicellular organisms, which proceeds through rapid and synchronous 
cleavage divisions that subdivide the fertilized egg into a large population of nucleated cells~\cite{kane1993zebrafish,gilbert2000introduction}. 
For instance, a frog egg, divides into 37,000 cells in $\sim$ 43 hours, while the Drosophila embryo proliferates to 50,000 cells in about 12 hours~\cite{gilbert2000introduction}. The early development of 
many animals is characterized by fast, synchronized divisions that occur are almost independent on inter cell interactions. This is followed by the midblastula transition accompanied by the loss of  synchrony in cell division and a lengthening of the cell cycle~\cite{kane1993zebrafish, kimmel1995stages}. 

The observation of spontaneous synchronization dates back to the 17${th}$ century when Christiaan Hyugens reported the synchronization of pendulums, due to transfer of energy from one to another. However, the reverse process of how  asynchronous behavior emerges in a synchronous system is a fascinating question, with relevance to understanding morphogenesis in living systems.  
Given that cell collectives generate geometric order, driven predominantly by cell divisions and traction forces~\cite{gibson2006emergence}, a better understanding the  transition between 
synchronous to asynchronous cell division during early development of many animals could provide important insights into this process. Based on the analysis of the experimental data, coupled with theory and simulations, we show that the synchronous to asynchronous transition (SAT), during zebrafish morphogenesis, provides vital clues about how cells integrate biochemical and mechanical signals in regulating cell division.  

Regulation of cell division is necessary for robust 
tissue growth and morphogenesis, with rapid cell 
proliferation being 
an integral phenotype characterizing early development~\cite{thompson1942growth,shaw2009wound}. In cells that undergo cleavage divisions, maturation promoting factor (MPF) undergoes cyclical changes that drive mitosis~\cite{gerhart1984cell,gilbert2000introduction}. Biochemically, the cyclical activity of MPF is driven by the protein kinase, cyclin B, that accumulates during the DNA synthesis phase of the cell cycle, which is degraded once the cells reach the mitotic 
phase~\cite{evans1983cyclin,swenson1986clam,gilbert2000introduction}. In combination with the biochemical factors that regulate the cell behavior, changes in the embryonic physical properties can feed back onto biochemical dynamics~\cite{samarage2015cortical,maitre2016asymmetric,gross2017active}. In particular, mechanical pressure, built up due to rapid cell divisions may function as a feedback signal regulating cell division~\cite{shraiman2005mechanical}. Here, we 
discover that the SAT transition commonly observed during embryo development serves as a powerful example in  dissecting the contribution of mechanical forces in controlling cell division. By analyzing experimental data, and developing theory and simulations of 3D multicellular collectives, we show that pressure experienced by cells due to mechanical interaction with their neighbors {\it quantitatively} explains the onset of the SAT. Furthermore, we predict that the lengthening of the cell cycle is a consequence of mechanical feedback on cell division.  

\begin{figure}
\centering
\includegraphics[width=0.98\linewidth]{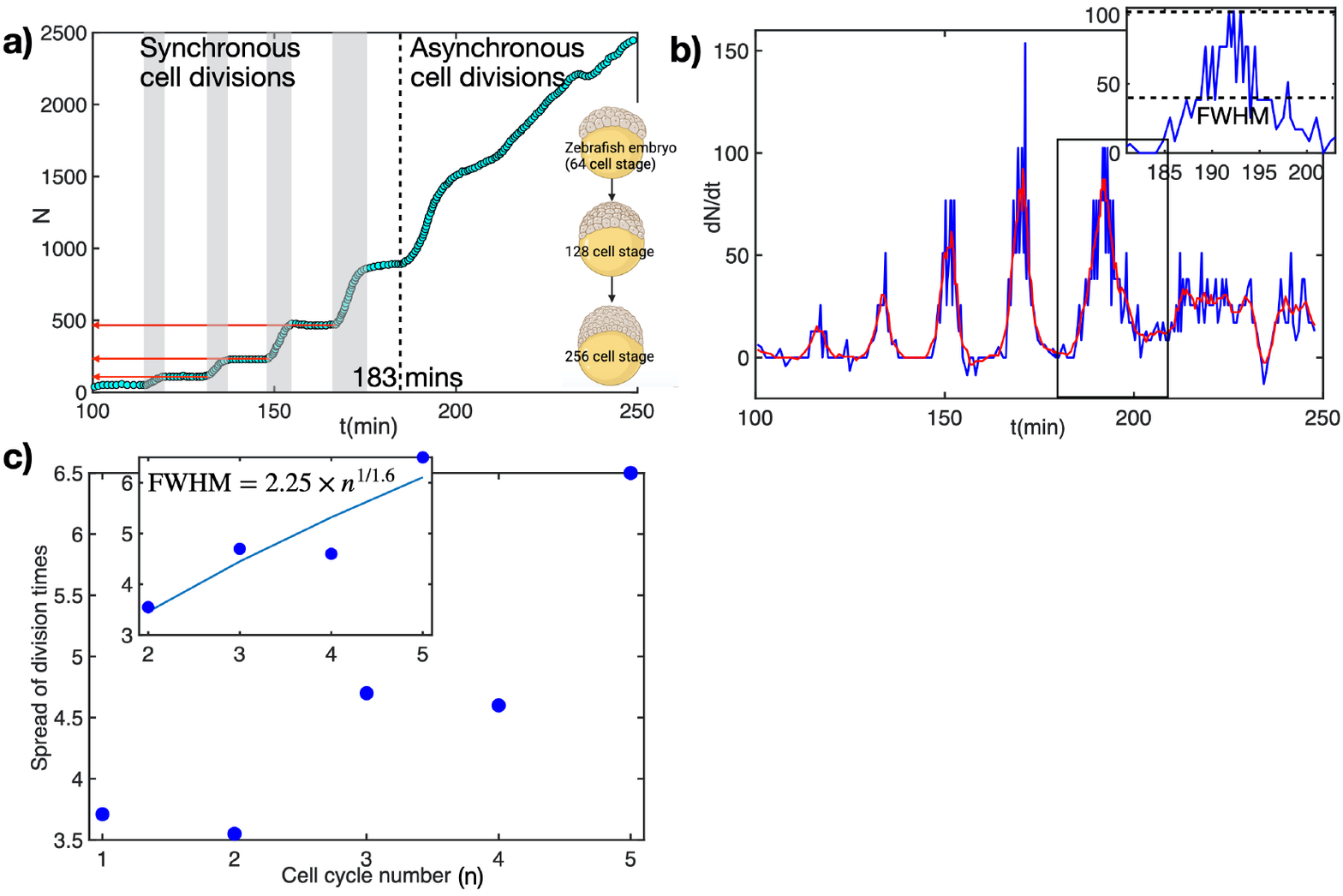}
\vspace{-2mm}
\caption{\footnotesize \textbf{Synchronous to asynchronous transition in cell division during embryo development} 
\textbf{(a)} Experimental data for the number of nuclei in the animal hemisphere of the Zebrafish embryo from the 64-cell stage to 2,000+ cells.  
Initially, cell divisions are highly synchronized (shaded portion) \textcolor{black}{exhibiting staircase-like pattern in the time dependence of cell number growth. After roughly five cell division rounds, the asynchronous behavior 
sets in (at times past the dashed line), where the cell number increases essentially continuously, without the staircase pattern observed at early times.} 
{\bf (b)} The time derivative of $N(t)$ (growth rate)  exhibits an oscillatory pattern where peaks indicate rapid cell division events and valleys represent time points where divisions do not occur. 
Peak widths, shown in the inset, is a measure of the heterogeneity in the cell cycle times. 
\textbf{(c)} 
Heterogeneity in the cell cycle time extracted from the full width at half maximum (FWHM) of the peaks associated with division events in {\bf (b)}. The FWHM increases as a function of the cell cycle number. The inset shows the fit to the data. 
}
\label{fig1:intro}
\end{figure}

\textit{Experimental Motivation:}
Our study is motivated by experimental data on time-dependent pattern of cell division during the early stages of Zebrafish morphogenesis. 
We start with a brief description of the experiment \cite{keller2008reconstruction}, which tracks the number of cell nuclei (a proxy for the number of cells), $N(t)$, as a function of time ($t$) using high-speed laser light-sheet fluorescence microscopy. Fig. \ref{fig1:intro}a shows a plot of $N(t)$ as a function of $t$ observed in the experiments. The cell divisions occur on the animal pole of the Zebrafish embryo (see inset showing three distinct division cycles), roughly every  $\sim 
20~\mathrm{mins}$~\cite{keller2008reconstruction,kane1993zebrafish}. 

We focus on two distinctive features of the data. First, in the initial regime ($100 ~min \leq t \leq 185~min$), $N(t)$ exhibits a clear step-like pattern (see arrows in Fig~\ref{fig1:intro}a).  Time intervals, with constant $N$, are interspersed with time regimes where the cell number increases (see the shaded regions). 
The step like pattern is a consequence of synchronous cell division, which is punctuated by finite time regimes where there are no cell division events.  
Second, at later times ($t\geq 185~min$), the staircase pattern in $N(t)$ disappears, signaling the onset of the synchronous to asynchronous transition (SAT), which occurs at $n^{*}_{E} \sim 5$ rounds of cell division. 
Here, the subscript ``E'' denotes experiments. These interesting features of cell division raise the following questions: what sets the time scale at which the SAT transition occurs? and what, if any role, does biomechanical feedback  plays on cell division in the SAT transition? These questions  motivated us to first develop the theoretical framework.

\begin{figure}
\centering
\includegraphics[width=0.99\linewidth]{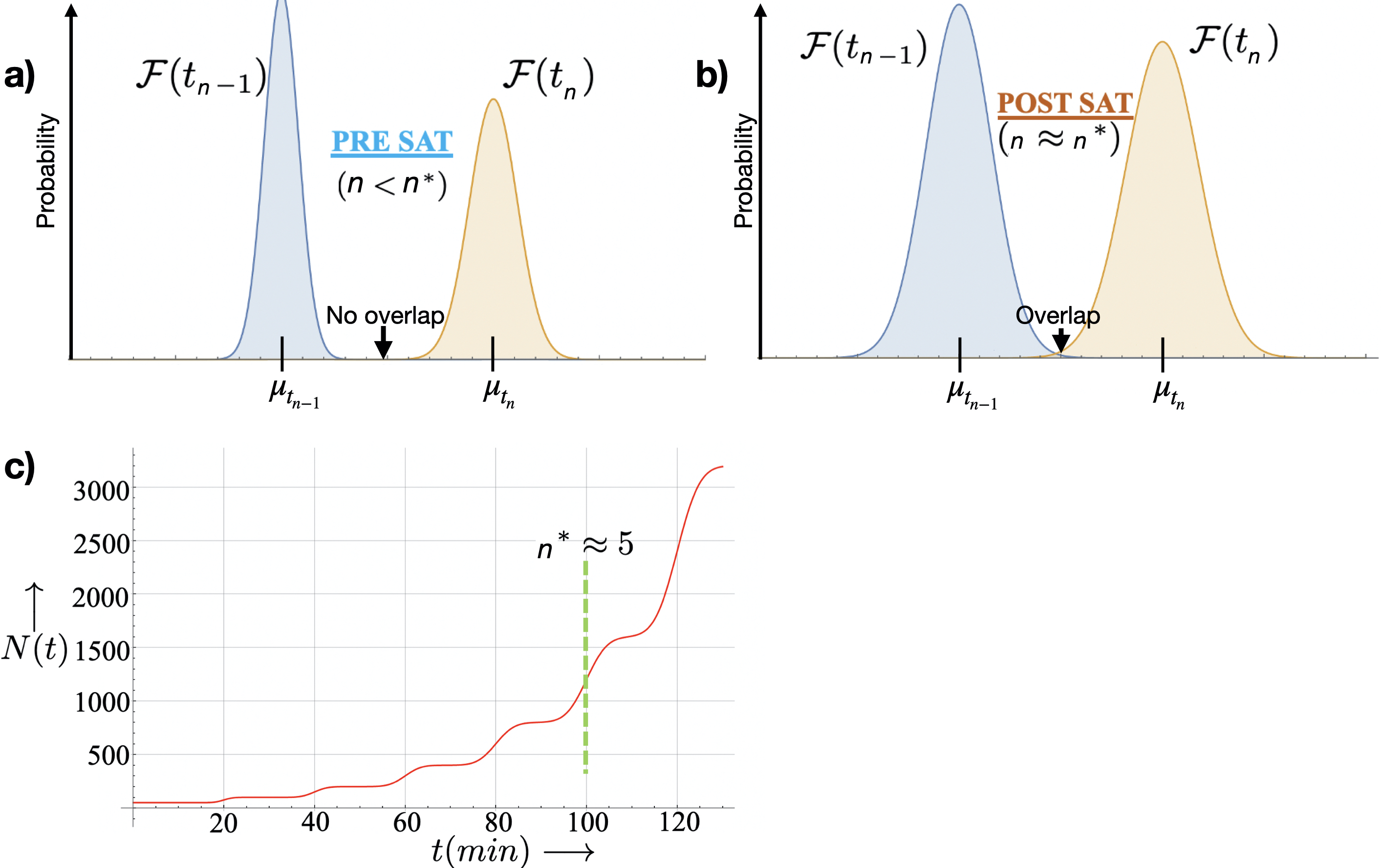}
\vspace{-5mm}
\caption{{\bf Variation in cell division times}. 
\textbf{(a)} Criteria used in the theory to predict the onset of the SAT in cell division. In the pre-SAT regime, the 
probability distributions, $\mathcal{F}(t_{n-1})$ and $\mathcal{F}(t_{n})$ of the times at which cells divide, do not overlap, which implies that cells divide independently. 
\textbf{(b)} In the post-SAT regime, we propose that the distributions, $\mathcal{F}(t_{n-1})$ and $\mathcal{F}(t_{n})$, begin to overlap, signaling the importance of cell-cell interactions. 
{\bf (c)} Staircase-like pattern in $N$ vs $t$ curve  calculated using theory. The plot shows 
the number of cells, $N(t)$, as a function of time (t) for multiple cell division rounds ($n$). 
The green dashed line demarcates the time  at which SAT occurs resulting in $n^{*}\approx 5$. We reproduce the experimentally observed $N(t)$ vs $t$ curves in Zebrafish Morphogenesis and the number of cell division cycles necessary for the onset of SAT~\cite{kimmel1995stages}.
}
\label{theory_scheme2}
\end{figure}

{\it {Increasing fluctuations in cell division times drive the SAT:} } The experimental data~\cite{keller2008reconstruction} in Fig.~\ref{fig1:intro}a makes clear that the SAT occurs after $n^{*}_{E} \sim 5$ (the subscript $E$ denotes experiments) rounds of cell division. Our objective is to provide a theory for predict the number of rounds, $n^{*}_{T}$ ($T$ stands for theory), it takes for the Zebrafish to exhibit SAT. 

Let the $n^{th}$ round of division for an individual cell occur at time, $t_n$. Assuming that $t_n$ is a random variable, we  write it as, 
\begin{equation}
    t_n=\sum_{i=1}^{n}\tau_i,
    \label{sum_tau}
\end{equation}
where $\tau_i$ is the cell cycle time for the $i^{th}$ round of cell division (see Fig. 2 in the SI). We assume that all the $\tau_i$'s are 
independent and identically distributed random numbers with the 
statistics of $\tau_i$'s being  given by $\tilde{\mathcal{P}}(\tau)$ as derived in section I in the SI.
From Eqn.(\ref{sum_tau}), we can readily obtain the statistics of $t_n$. We will focus on the mean ($\mu_{t_n}$) and the variance ($\sigma^2_{t_n}$) because they can be compared to the experimental data \cite{puliafito2012collective, keller2008reconstruction}. The mean, $\mu_{t_n}$, is given by
\begin{equation}
    \mu_{t_n}=\sum_{i=1}^{n}\mu_{\tau,i}=n\mu_{\tau},
    \label{mean_tp}
\end{equation}
and the variance, $\sigma_{t_n}^2$, is given by
\begin{equation}
    \sigma_{t_n}^2=\sum_{i=1}^{n}\sigma_{\tau,i}^2=n\sigma_{\tau}^2.
    \label{var_tp}
\end{equation}
Here, $\mu_{\tau}$ and $\sigma_{\tau}$ are the mean and standard deviation of a single cell cycle round (see SI section I for details). As can be seen from Eqs. (\ref{mean_tp}) and (\ref{var_tp}), both the mean and the fluctuations (variance) increase linearly with the number of cell cycle rounds, $n$. Also, we predict that the probability distribution for $t_n$, $\mathcal{F}(t_n)$ given by
\begin{equation}
 \mathcal{F}(t_n)=\mathcal{N}(t_n; n\mu_\tau, \sqrt{n}\sigma_\tau), 
 \label{tp_prob_den}
\end{equation}
\textcolor{black}{where $\mathcal{N}$ is a normal distribution with mean $n\mu_\tau$, and standard deviation $\sqrt{n}\sigma_\tau$}. 

To test the theoretical predictions, we  quantified the statistics of cell cycle times from the experimental data. If our model is correct, we expect that the fluctuations in cell division times should increase at later cell division cycles. 
Remarkably, as predicted by theory (see Fig.~\ref{fig1:intro}) 
the variance increases as $n$ increases. In Fig. \ref{fig1:intro}b, we plot $\frac{dN}{dt}$ versus $t$, which allows us to extract the spread in cell division times for different $n$ from the full width at half maximum (FWHM) of each peak (see Fig.~\ref{fig1:intro}b and inset). 
The spread in cell division times scales as 
$\sigma_{t_p}\propto n^{\frac{1}{1.6}}$, which is close to the theoretical prediction ($\sigma_{t_p}\propto n^{\frac{1}{2}}$). 
The modest deviation from theory might be due to poor statistics, suggesting that robust experiments should be performed to test the functional dependence of fluctuations with time. 
We note that Eq. (\ref{tp_prob_den}) can be interpreted as a particle undergoing Brownian motion advected by a constant ``velocity'' $\mu_{\tau}$. Variability in cell cycle duration has been observed  long ago, but its origin is  unclear~\cite{smith1973cells, sandler2015lineage}. Our theory suggests that it is driven by fluctuations in cell division times. 
  
With the ansatz for $\mathcal{F}(t_{n})$, we can estimate $n^{*}$ where SAT occurs. The onset of 
synchronous to asynchronous 
transition (SAT) corresponds to the absence of a plateau in the $N$ as a function of  $t$ curves between successive cell division epochs. This is realized when the fluctuations in $t_{n^{*}}$ are so large that the cell division time probabilities at two consecutive cell cycles - $\mathcal{F}(t_{n^{*}-1})$ and $\mathcal{F}(t_{n^{*}})$ - begin to overlap (see Figure \ref{theory_scheme2}a and \ref{theory_scheme2}b). Mathematically, this is equivalent to writing,  
\begin{equation}
\mu_{\tau} \approx \lambda \big[\sigma_{t_{n^{*}-1}}+\sigma_{t_{n^{*}}}\big], 
\label{approx}
\end{equation}
where $\lambda$ controls the strength of the overlap between two successive distributions, $\mathcal{F}$'s. 
We assume  that the two distributions start to overlap at three standard deviations from the mean, $\lambda=3$ (see Fig.~\ref{theory_scheme2}b). Note that $\lambda$ can vary  depending on the strength of the overlap between the cell division time probability distributions. Using $\lambda=3$, we can  calculate $n^{*}$, the cell cycle round when SAT occurs given by,
\begin{equation}
    n^{*}=\bigg(\frac{\kappa^2+1}{2\kappa}\bigg)^2,
    \label{pred}
\end{equation}
where $\kappa=\frac{\mu_{\tau}}{3\sigma_{\tau}}$. In the experiments, $\sigma_{\tau}=1.6~min$ and $\mu_{\tau}=20~min$. 
Therefore, $n^{*}=4.85\approx 5$ from Eqn. (\ref{pred}) (refer to Figure 3 in SI for dependence of $n^{*}$ on $\lambda$),  close to the experimental data in Figure \ref{fig1:intro}a, where the transition occurs after five cell division rounds. The theory also predicts that at the transition, $\sigma_{t_{n^{*}}}=\sqrt{5}\sigma_{t_{n=1}}=3.6~min$, which is close to the experimentally observed value of  $2.8~min$. The choice of $\lambda=3$ is justified {\it a posteriori}.

\begin{figure}
\centering
\includegraphics[width=0.8\linewidth]{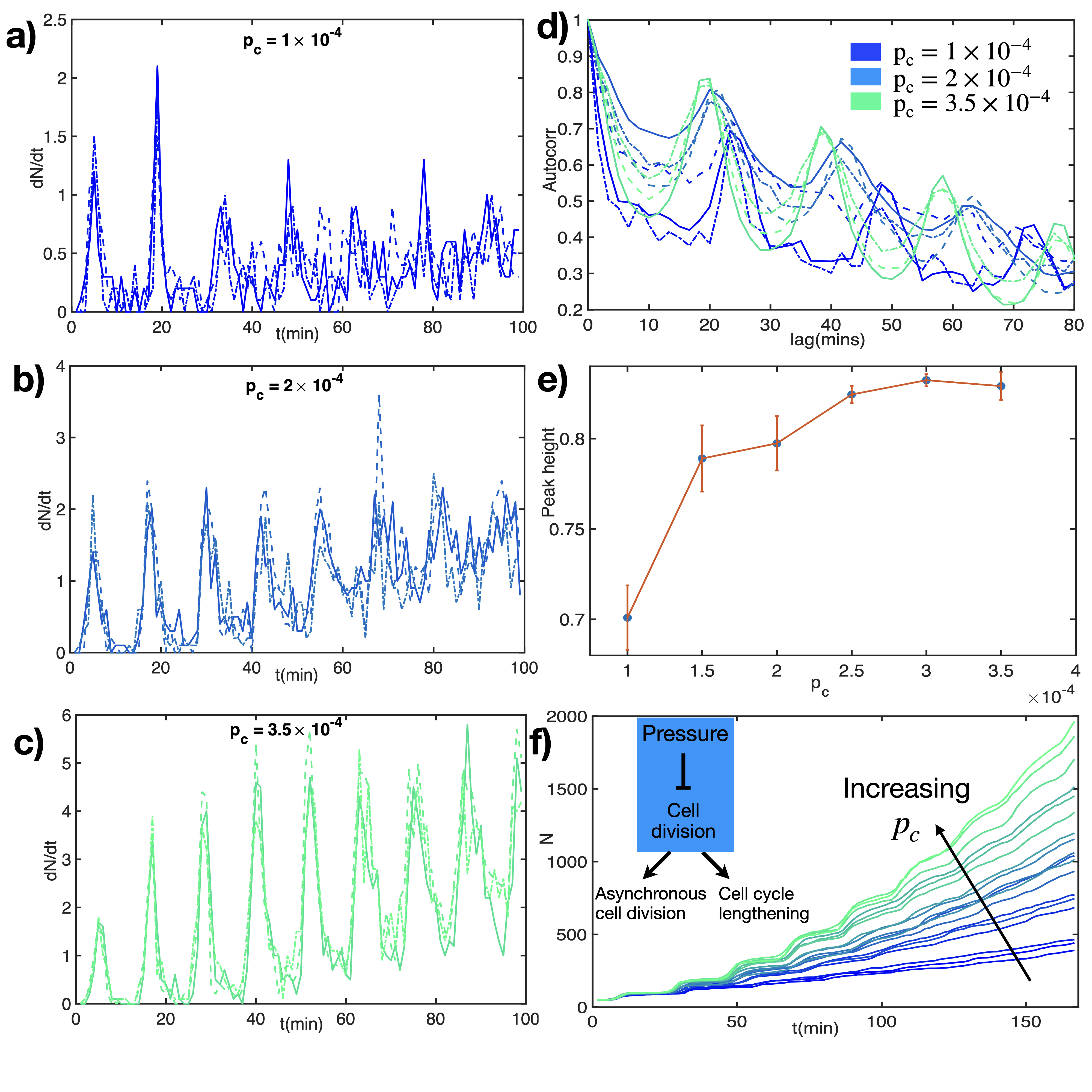}
\vspace{-1mm}
\caption{\textbf{Mechanical feedback determines synchronous cell division}  
\textbf{(a)} Division rate as a function of time at low $p_c = 1\times 10^{-4} MPa$. Low $p_c$ value corresponds to heightened cell sensitivity to mechanical feedback on cell growth and division.  
\textbf{(b-c)} Division rate as a function of time at intermediate and high  $p_c = 2.5\times 10^{-4} MPa$ and $p_c = 3.5\times 10^{-4} MPa$, respectively. 
Higher $p_c$ value, corresponding to lesser cell sensitivity to mechanical feedback on cell growth and division, leads to highly synchronous cell division for a longer time range. 
\textbf{(d)} Autocorrelation of cell division rate at 3 $p_c$ values. The height and time point of the first peak  determine the strength of periodicity and the interval between repeated peaks in cell divisions. 
\textbf{(e)} Peak height of the autocorrelation function at varying sensitivity to mechanical feedback through pressure. Strength of periodic division rate increases with $p_c$ and saturates at higher values. 
\textbf{(f)} Number of cells vs time as 
generated from simulations incorporating mechanical feedback on cell divisions. Inset: Schematic 
illustrating how increasing pressure limits cell divisions, leading to cell cycle lengthening and asynchronous cell divisions.  
}
\label{theory_scheme3}
\end{figure}
{\it Step-like pattern in cell number growth is recapitulated in theory}:
The theory also captures the step-like pattern in the growth of $N(t)$
observed in experiments (see Figure \ref{fig1:intro}a). We  estimate that $N(n,t)$ which denotes the number of cells at 
time $t$ after $n$ rounds of cell division from the following relation, 
\begin{equation}
    N(n,t)=N(n-1,t)\big[1+\xi(n,t)\big],
    \label{no_cell}
\end{equation}
where $\xi(n, t)$ is the probability that a cell has undergone $n$ rounds of cell division until time $t$. In particular, $\xi(n, t)$ is the cumulative distribution function (CDF) associated with $\mathcal{F}(t_{n})$  is given by, 
\begin{equation}
    \xi(n,t)=\int_{-\infty}^{t}\mathcal{N}[s;\mu_{t_n},\sigma_{t_{n}}]ds=\frac{1}{2}[1+erf\bigg(\frac{t-\mu_{t_n}}{\sqrt{2} \sigma_{t_{n}}}\bigg)],
    \label{prob_n}
\end{equation}
where $erf$ is the error function. In the above equation, the lower integration limit is from $-\infty$ because the support of the normal distribution is from $[-\infty, \infty]$. However, since the standard deviation, $\sigma_{t_{n}}$,  is  small compared to $\mu_{t_n}$, the majority of the contribution to integral in Eq. \ref{prob_n} arises from $[\mu_{t_n}-3\sigma_{t_{n}}, \mu_{t_n}+3\sigma_{t_{n}}]$. Substituting the expression for $\xi(n, t)$ into Eq. \ref{no_cell}, we obtain the 
 recursive relation,
\begin{equation}
    N(n, t)=N(n-1, t)+\frac{N(n-1,t)}{2}\bigg[1+erf\bigg(\frac{t-\mu_{t_n}}{\sqrt{2} \sigma_{t_{n}}}\bigg)\bigg].
    \label{n_vs_t}
\end{equation}
Since the initial condition, $N(n=0, t=0)$ is known, we can calculate $N(n,t)$ for any $n$ and $t$. Remarkably, Figure \ref{theory_scheme2}c, which shows the plot for $N(n,t)$ as a function of $t$  using Eqn. (\ref{n_vs_t}), 
reproduces the step-like pattern observed during 
Zebrafish morphogenesis (compare to Figure \ref{fig1:intro}a and Figure \ref{theory_scheme2}c). 

An essential question in cell biology is whether cell cycle control is deterministic or stochastic~\cite{nurse1980cell}, and whether the observed variability in division times originate mainly from stochasticity, or from deterministic processes that appear to be random~\cite{shields1977transition, sandler2015lineage}. 
Although the theory shows that synchronous cell division may be quantitatively described without considering interactions between cells,  the onset of asynchronous cell division, with increasing cell numbers, points to the importance of the local physical forces experienced by cells upon division~\cite{shraiman2005mechanical, irvine2017mechanical, malmi2022adhesion}. Indeed, seminal studies have demonstrated that variations in cell density  control cell proliferation rate through contact inhibition of proliferation~\cite{godard2019cell}. Contact inhibition is thought to restrict cell proliferation when space is limited with dense spatial arrangements of cells leading to strong suppression of cell division~\cite{eagle1967growth, pavel2018contact}. To determine whether such mechanical effects  influence SAT, we turn to a three-dimensional (3D) computational model for cell divisions.  

{\it Biomechanical feedback controls the SAT:}
Cells that are constantly in contact with their neighbors experience  compressive forces that limit their ability to progress through the cell  cycle. The cumulative effect of such compressive forces from the cell microenvironment can  characterized at the multicellular scale through pressure experienced by the cells. Such cell-cell contact-induced pressure is expected to decrease the rate of division~\cite{shraiman2005mechanical,puliafito2012collective, malmi2022adhesion}. The possibility of mechanical feedback on cell division has significant  implications, because cell cycle regulation is important in stem cell biology as pluripotent cells divide slowly over long periods but can initiate growth on demand. 
To capture these effects, we performed particle-based 
 simulations~\cite{malmi2018cell, sinha2020self, sinha2021inter, sinha2023observe, malmi2022adhesion, marchant2022cell, zills2023enhanced},  to investigate how cell cycle progression, controlled via feedback based on  
local inter-cellular mechanical interactions, impacts the ability of  cells to grow and divide. 
The cell cycle time during each round of cell division is $\mu_{\tau}$. Cells may divide only if 
pressure experienced by a cell is below a critical pressure ($p<p_c$). This implies that synchronous divisions occur during  zebrafish morphogenesis before t mechanical feedback is operative. The universality of SAT can be 
tested by conducting high temporal resolution cell cycle experiments on different cell types~\cite{sandler2015lineage, cermak2016high}.

\begin{figure}
\centering
\includegraphics[width=1.0\linewidth]{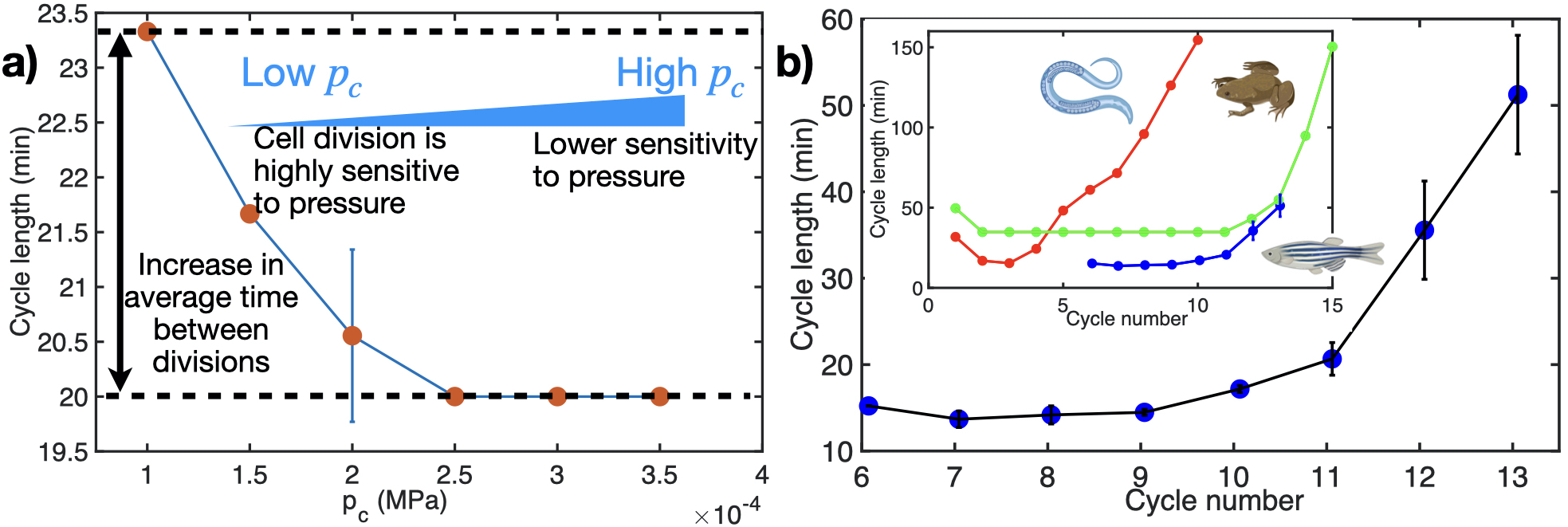}
\vspace{-5mm}
\caption{\textbf{Mechanical feedback elongates cell cycle} {\bf (a)} Average cell cycle length increases with stronger mechanical feedback. Cell cycle length is quantified from the time of the first peak in the growth rate autocorrelation function.  
{\bf (b)} Experimental data extracted from Ref.~\cite{kane1993zebrafish} for cell cycle length during  cycle numbers 6-13 for Zebrafish development. Inset shows similar behavior that is conserved across multiple species - {\it C. elegans} (roundworm), {\it X. laevis} (frog) in addition to {\it D. rerio} (zebrafish). 
} 
\label{fig4}
\end{figure}

In our model, cells are either in the dormant (D) or in
the growth (G) phase. Even though cells in the experiment  do not  grow, we consider physical cell growth as a proxy for cell cycle progression. 
We track the sum of the normal
pressure, $p_i$ that a particular cell $i$ experiences due to contact
with its neighbors, using 
\begin{equation}
    p_i = \Sigma_{j \in NN(i)} \frac{|\vec{F}_{ij}\cdot\vec{n}_{ij}|}{A_{ij}}, 
\end{equation}
where $\vec{F}_{ij}$ is the force that $i^{th}$ cell experiences from its nearest neighbors ($NN$) $j$, $\vec{n}_{ij}$ is the unit vector that points from the center of cell $j$ to the center of cell $i$, and the contact surface area between cells $i$ and $j$ is $A_{ij}=\pi h_{ij} R_i R_j / (R_i + R_j)$. The nearest neighbors of a cell is determined through  
positive overlap between cells through the  $h_{ij}$, defined as $\mathrm{max}[0,R_i + R_j - |\vec{r}_i - \vec{r}_j|]$. Importantly, once the pressure exceeds the critical pressure $p_c$ the cell stops growing and enters  a dormant phase where cell division cannot occur. 
Lower $p_c$ implies that cells enter the dormant phase with high probability while higher $p_c$ values would make the cells less sensitive to feedback on division due to mechanical 
interactions with its neighbors. By varying $p_c$, we predict that higher 
sensitivity to mechanical feedback on cell growth leads to faster onset 
of asynchronous cell divisions. The growth rate of the cell collective closely recapitulate experimental data 
where clear peaks with bursts of cell division events followed by time 
regimes with absence of division events 
(see Figs.~\ref{theory_scheme3}a-c). 
Notably, the synchronous division events are more persistent 
at higher $p_c$ and are maintained for longer times as compared to lower $p_c$ where the clear peaks in division rate are quickly washed out (compare Figs.~\ref{theory_scheme3}a and c). 

To quantify the synchronous nature of the cell division events, we evaluated 
the autocorrelation of the division rate to verify the presence of 
cycles and determine their duration. The autocorrelation  function of a periodic signal has the  same periodic characteristics,  evaluated using, 
\begin{equation}
C(t_{\gamma}) = \langle \Sigma_{\alpha,\beta}  \frac{dN(t_{\alpha})}{dt}\times  \frac{dN(t_{\beta})}{dt} \delta (t_{\gamma} - (t_{\alpha} - t_{\beta}))\rangle,
\label{autocorr}
\end{equation}
where, $t_{\gamma}$ is the lag time between time points $t_{\alpha},t_{\beta}$, and $\delta(z)=1$ if $z=0$ and $0$ otherwise. The autocorrelation function is normalized to ensure that $C(t_{\gamma}=0)=1$. 
We focus on the first peak in the autocorrelation function in order to quantify the peak height and peak location as a function of $p_c$. The peak height measures the strength of the oscillatory division patterns. 
The peak height increases and saturates at high $p_c$, as shown in Fig.~\ref{theory_scheme3}e, which is indicative of more persistent synchronous division patterns with weak mechanical feedback. 
 (see Fig.~\ref{theory_scheme3}f).

\textit{Mechanical feedback lengthens the cell cycle:}
So far we have focused on the collective effect of coordination of cell division events between different cells.  We now delve into the single cell scale to verify our prediction that cell cycle lengthening is due to mechanical feedback. Many embryos exhibit a period of rapid cell division followed by a lengthening of cell division times during early development~\cite{farrell2014egg}. The embryo developmental processes are crucially linked to the speed of cell division as well as the cell cytoskeleton during mitosis 
is engaged in building the mitotic spindle~\cite{o2004metazoans,farrell2014egg}. As many of the morphogenetic events require cytoskeletal arrangements, they must wait for cell divisions to slow down~\cite{farrell2014egg}. As such slowing of the cell division time, associated with the mid blastula transition, is an important and conserved feature of embryo development. 

We anticipate that pressure-dependent feedback on cell division may lead to the lengthening of the cell cycle.  Recently, we showed that pressure experienced by cells near the core of multicellular spheroids is  elevated compared to those at the periphery, setting up  emergent variations in the cell division rate between the core and periphery~\cite{malmi2022adhesion,sinha2020spatially,sinha2020self}. Moreover, there is a
the coupling between cell growth rate and pressure through intracellular molecular crowding was~\cite{alric2022macromolecular}. 
From the autocorrelation in cell growth rate (see Eq.~\ref{autocorr}), the lag time at which the peak is located is a measure of the time difference between two division events, 
which corresponds to the average cell cycle time. Our simulations show that the cell cycle time is inversely proportional 
to the value of the threshold pressure, $p_c$ (see Fig~\ref{fig4}a). Lower values of $p_c$ correspond to higher sensitivity to cell division because  mechanical forces exerted by cells on one another could easily reach a lower value of $p_c$ (see Fig.~\ref{fig4}b inset). Our quantification shows that higher sensitivity to mechanical feedback lead to 
a longer cell division time. We next turned to the experimental 
data to assess whether cell cycle lengthening is observed during Zebrafish development. Indeed, as shown in Fig.~\ref{fig4}b, the cell cycle lengthens markedly as cells progress through repeated 
division cycles. Experimental data, therefore, shows that SAT is 
followed by an increase in the mean cell cycle \cite{kimmel1995stages} during Zebrafish morphogenesis. The data in this regime can be fit using $\mu_{\tau} (p, p_c)=\mu_{\tau}e^{\frac{p}{p_c}}$, where the exponential increase in cell cycle times is a consequence of mechanical feedback.

\textit{Conclusion}
The finding that synchronous cell division depends on the strength of mechanical feedback points to an interesting manifestation of the cross-talk between mechanical forces and cell cycle regulation during  embryo development. Here, we showed that coordination between cells during cell cycle progression is regulated by whether cells experience strong or weak mechanical feedback. A direct implication of our study is that suppression of mechanical feedback, which may be realized by partial knockout of E-Cadherin, would prolong the synchronous phase.
Given our prediction that asynchronized cell division emerge during the course of embryo development at the cusp of the mid-blastula transition, corresponding to the activation of transcription, SAT  in cell division could be a signal for cells to trigger their own transcription programs as a function of increasing cell density. Cells that grow under strong mechanical feedback, transition rapidly into asynchronous cell division patterns even at very values of small pressure. In contrast, for cells growing under weak mechanical feedback, synchronous division patterns persist for a longer period of time. 

We have shown that mechanical feedback drives the lengthening of the cell cycle time. Cells experiencing strong mechanical feedback elongate their cell cycle. On the other hand, cells under weak mechanical feedback progress through the cell cycle more rapidly and more uniformly. The proposed feedback mechanism on growth is therefore a `mechanical' intercellular signal that enables cells to measure density and the timing of development. 

The observation that embryos develop autonomously under a predetermined schedule is striking 
\cite{kageyama202225,perrimon2012signaling, artavanis1999notch}. 
Studies based on molecular biology revealed that various genetic factors control the periodicity of the development of somites in which a network of genes whose expression oscillates 
synchronously and regulate the timing of development. As tissue mechanics was recently shown to be an important factor in the formation of body segments, formerly thought to be based on segmentation clock~\cite{naganathan2022left}, we propose that mechanical feedback on cell growth  serves as a  clock (timing device)  that mediates asynchronization of cell division, and therefore, the onset of the transcriptional machinery during the mid-blastula transition.

\bibliographystyle{naturemag}
\bibliography{ms}
\end{document}


\title{Supplementary Information: On the role of mechanical feedback in synchronous to asynchronous transition during embryogenesis}

\author{Abdul N. Malmi-Kakkada}
\email{amalmikakkada@augusta.edu}
\affiliation{Department of Physics and Biophysics, Augusta University, Augusta, GA 30192, USA}

\author{Sumit Sinha}
\email{ssinha@g.harvard.edu}
\affiliation{Department of Physics, University of Texas, Austin, TX 78712, USA}
\affiliation{Applied Mathematics, School of Engineering and Applied Sciences, Harvard University, Cambridge, MA, USA}
\affiliation{Department of Data Science, Dana-Farber Cancer Institute, Boston, MA, USA}

\author{D. Thirumalai}
\email{dave.thirumalai@gmail.com}
\affiliation{Department of Physics, University of Texas, Austin, TX 78712, USA}
\affiliation{Department of Chemistry, University of Texas, Austin, TX 78712, USA}

\date{\today}

\pacs{}

\maketitle


\section{Distribution for single cell cycle division times }
The early embryonic cell cycles lack the layers of regulation that is present in more mature cells~\cite{siefert2015cell}.  
To understand the role of mechanical feedback on cell cycle regulation, we introduce a minimal model. The multiple cell cycle stages, shown in Fig.S1a, are represented as discrete steps.   Once a cell goes over all the 
stages, a cell undergoes division. Assuming that there are $n$ stages,  let us denote the $i^{th}$ stage of a cell as $G_{i}$. We denote transition rate from $G_{n-1} \rightarrow G_{n}$ as $k_{n}$. We assume, for simplicity, that $k_1=k_2=....k_{n-1}=k$. The functional dependence of $k$ on mechanical feedback ($p$) is given by,
\begin{equation}
k  = \left\{
        \begin{array}{ll}
            k &  p \leq p_c \\
            ke^{-\frac{p}{p_c}} &  p > p_c
        \end{array}
    \right\},
\end{equation}
where $p$ is local mechanical stress on a cell, and $p_c$ is the stress threshold value above which the cell is dormant, at least temporarily (see Fig. S1b). The functional form of $k$ is inspired from experiments, where the cell division time increases as a function of tissue growth \cite{puliafito2012collective,kimmel1995stages}. 
For our  model, the time ($\tau$) it takes for a single cell to divide is given by,
\begin{equation}
    \tau=\sum_{i=1}^{n-1}t_i,
    \label{sum_ti}
\end{equation}
where $t_i$ is the  transition time from $G_{i-1}$ to $G_i$. Because the $t_i$'s are independent random variables, it follows that  $\tau$ is a random variable as well. The assumption that $k_i=k$ for all $i's$ implies that the probability density of $t_i$'s, $\mathcal{P}(t_i)$, are identical, $\mathcal{P}(t_i)=\mathcal{P}(t)$. The probability density, $\mathcal{P}(t)$, is governed by the differential equation, 
\begin{equation}
    \frac{d\mathcal{P}(t)}{dt}=-k\mathcal{P}(t),
\end{equation}
that is subject to the constraint $\int_{0}^{\infty}\mathcal{P}(t)dt=1$. The solution to the above equation is, 
\begin{equation}
    \mathcal{P}(t)=ke^{-kt}.
\end{equation}
By knowing the functional form of $\mathcal{P}(t)$, we can readily calculate all the moments of $t$. However, in this work, we  focus on the mean and the variance as they are the relevant observables in the experiment \cite{keller2008reconstruction}. The mean, $\mu_t=\langle t \rangle$, and the variance, $\sigma_t^2=\langle t^2 
\rangle - \langle t\rangle^2$, of $\mathcal{P}(t)$ are given as $\frac{1}{k}$ and $\frac{1}{k^2}$ respectively. Utilizing Eq. \ref{sum_ti}, we 
can calculate the mean ($\mu_{\tau}$) of the cell division time, $\tau$, which is given by
\begin{equation}
    \mu_{\tau}= \sum_{i=1}^{n}\mu_t= \frac{n}{k}, 
\end{equation}
and the variance ($\sigma^2_{\tau}$) of $\tau$ is
\begin{equation}
    \sigma_{\tau}^{2}=\sum_{i=1}^{n}\sigma_t^2=\frac{n}{k^2}.
\end{equation}
For zebrafish, $\mu_{\tau}\approx 20~min$ and $\sigma_{\tau}\approx 1.6~min$\cite{keller2008reconstruction}.

In order to analyze the experimental data, we fit the cell division times to a normal distribution (see Fig. 2a in the main text). Thus, for our analyses to be consistent with theory, we ought to recover the normal distribution from theory. This is possible in $n\rightarrow \infty$ limit, where we obtain $\tilde{\mathcal{P}}(\tau)=\mathcal{N}(\tau;\mu_\tau, \sigma_{\tau})$ utilizing the well known central limit theorem. Here, $\tilde{\mathcal{P}}(\tau)$ denotes the probability distribution of cell division times $\tau$, and $\mathcal{N}(\tau;\mu_\tau, \sigma_{\tau})$ is the normal distribution with mean $\mu_{\tau}$ and standard deviation $\sigma_{\tau}$. Note that $\tilde{\mathcal{P}}(\tau)$ is the distribution of time, $\tau$, for a cell to undergo a single cell division event. 

Physiologically, $n\rightarrow \infty$ implies that for a cell to divide, numerous physio-chemical events have to be completed, which occurs with astonishing regularity  in many organisms. 

\begin{figure}
\centering
\includegraphics[width=0.8\linewidth]{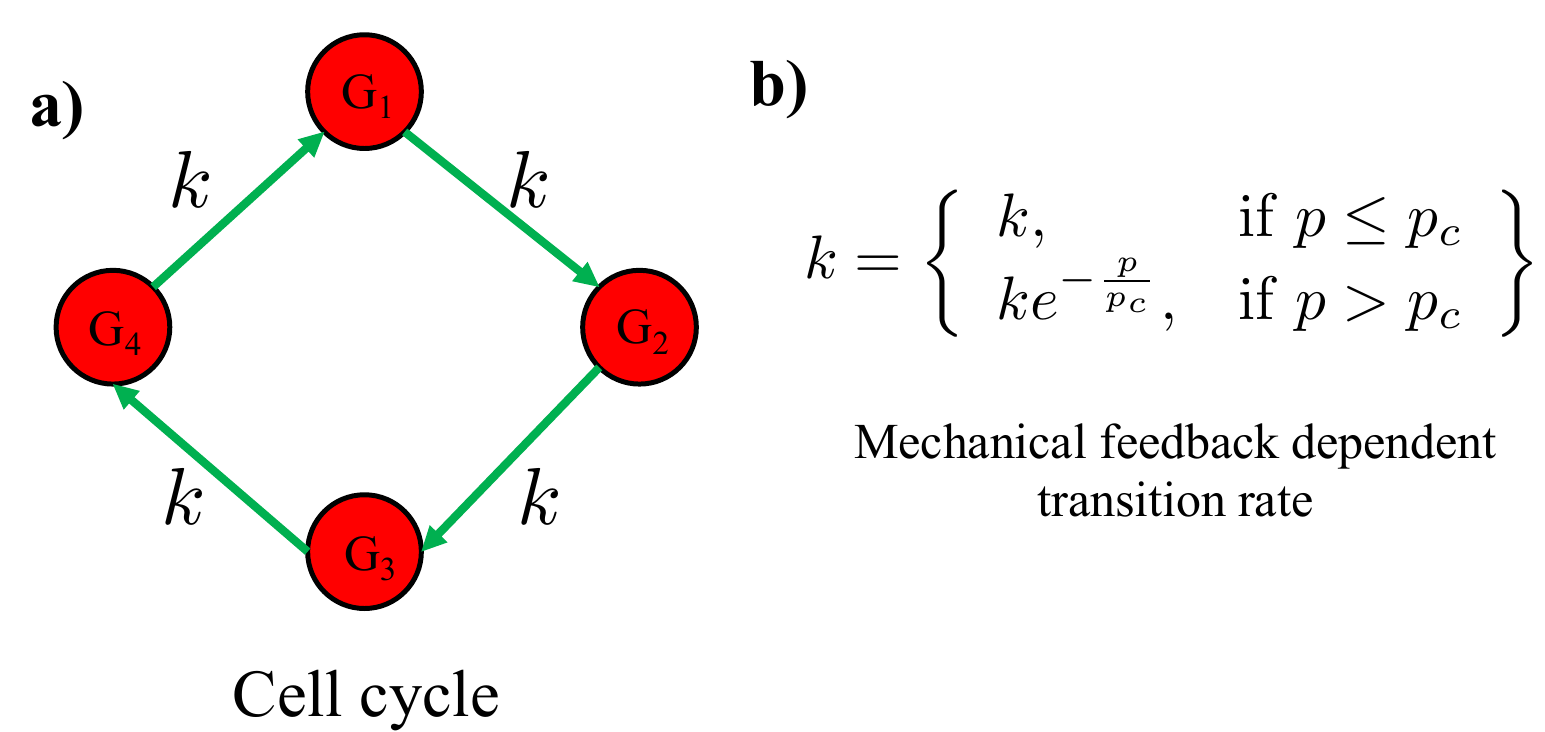}
\vspace{-5mm}
\caption{{\bf Schematic of stochastic model for cell division}. ({\bf a}) Cartoon depicting a cell cycle for an individual cell. In the present picture, the cell cycle is assumed to be comprised of $n=4$ stages ($G_1, G_2,G_3,G_4$). A single cell transitions from one stage to the other at the same rate, $k$. Once a cell has gone through all the stages, a cell is assumed to have undergone a single  cell division. (\textbf{b}) The transition rate between the stages depends on mechanical feedback which is controlled by the local environment (see the expression for $k$ given above.   }
\label{theory_scheme}
\end{figure}
\section{Choice of $\lambda$}
In Eq. 5 of the main text, we need the value of $\lambda$ to predict the transition point. The parameter $\lambda$  sets the degree of overlap between two cell division time probability distributions - $\mathcal{F}(t_{n^{*}-1})$ and $\mathcal{F}(t_{n^{*}})$. 
Figure \ref{n_lambda} shows the plot of $n^{*}$ as a function of $\lambda$. Approximately around $\lambda=3$, we get the right fit between  experiments and theory (see the main text). 

\begin{figure}
\centering
\includegraphics[width=0.8\linewidth]{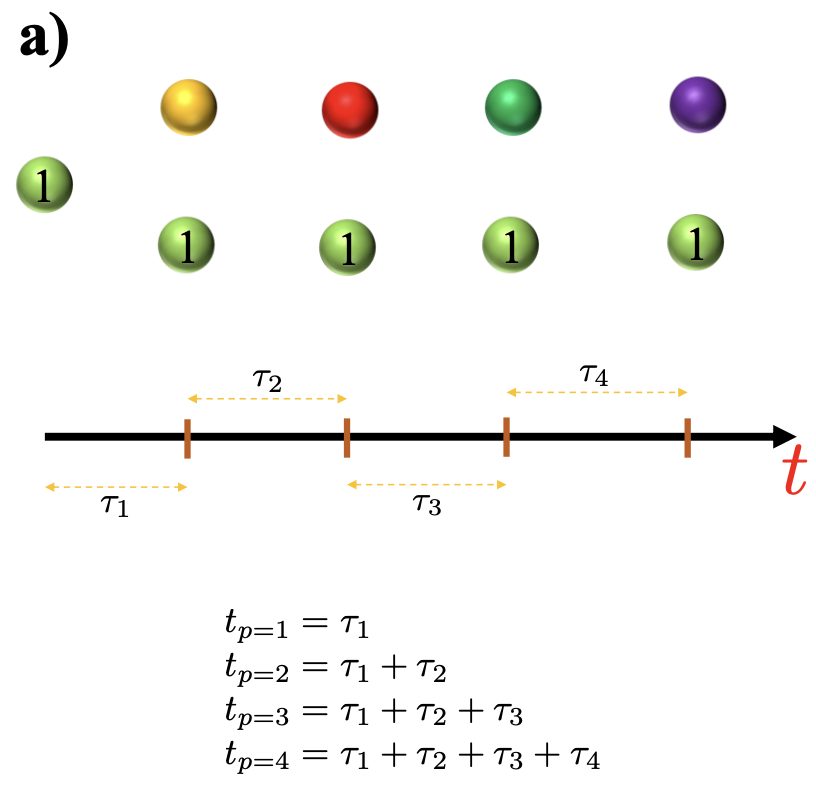}
\vspace{-5mm}
\caption{{\bf Schematic of stochastic model for cell division}. 
The schematic for a cell (in green labelled as $1$) undergoing subsequent cell divisions. The different colored cells at different time points are daughter cells. In the middle, a schematic for Eqn. 1 (main text) is shown. In the bottom, Eq. (1) (main text) is shown mathematically for different $t_p$, where $p$ is the number of cell division rounds. }
\label{theory_scheme}
\end{figure}

\begin{figure}
\centering
\includegraphics[width=0.8\linewidth]{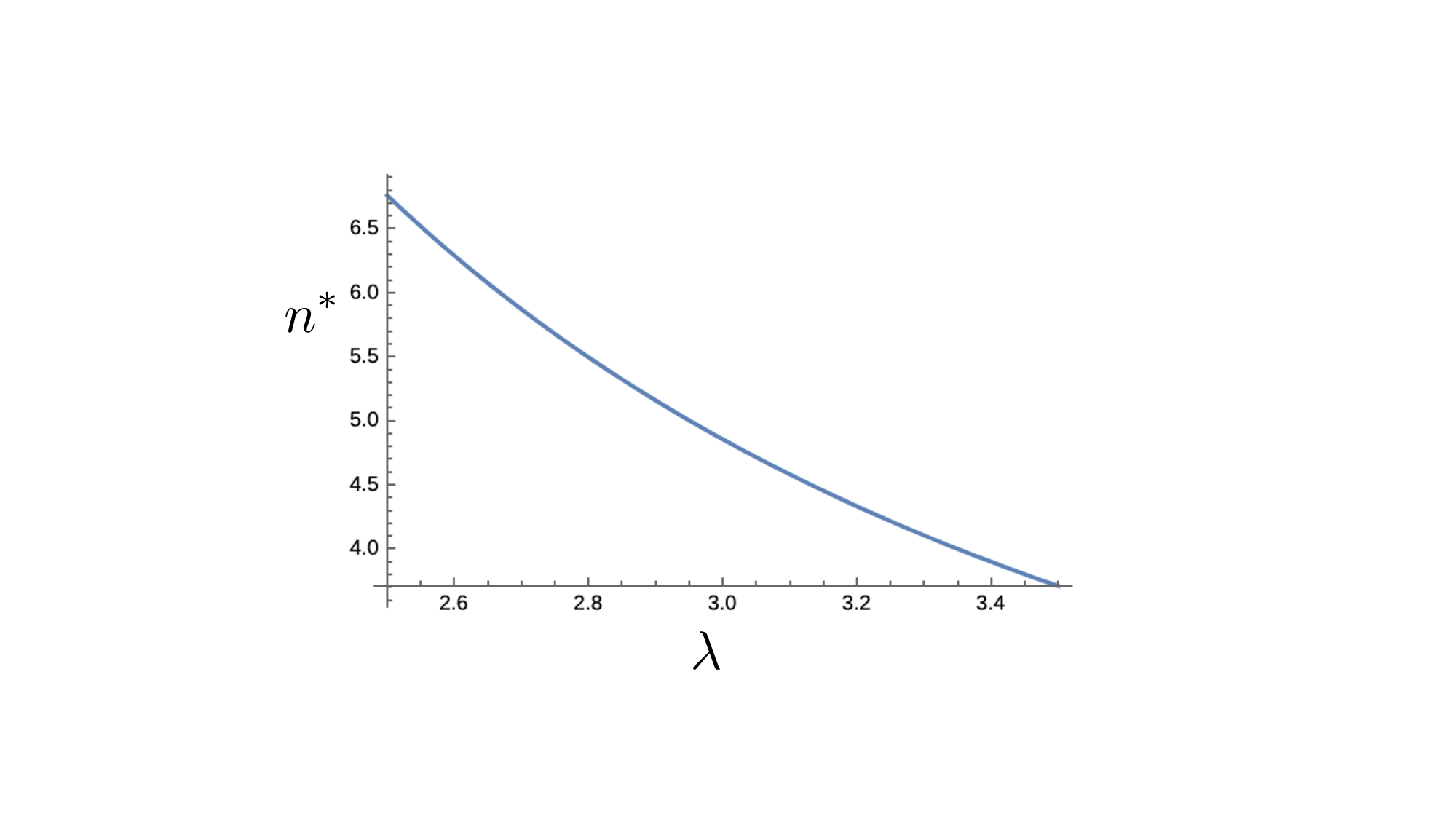}
\vspace{-5mm}
\caption{Plot showing the dependence of $n^{*}$ on $\lambda$. }
\label{n_lambda}
\end{figure}

\section{Simulation Details}
{\bf Initial Conditions:}
At time, $t=0$, we begin with 50 cells. The radii of the cells are sampled from a Gaussian distribution, 
$p(R_{i})=\frac{1}{0.5\sqrt{2\pi}} e^{-(R_{i}-4.5)^2/2\times 0.5^2}$. Similarly, the elastic moduli, $E_{i}$ and the Poisson ratio $\nu_{i}$ are obtained from a 
Gaussian distribution with standard deviation of $10^{-4}\mathrm{MPa}$ and $0.02$ respectively (the mean values are
given in Table 1).   
The receptor and ligand concentration on the cell surface are Gaussian distributed 
($p(c_{i}^{rec}/c_{i}^{lig})= \frac{1}{0.02\sqrt{2\pi}} e^{-(c_{i}^{rec}/c_{i}^{lig}-0.9)^2/2\times 0.02^2}$), centered around the mean (=0.9) with a dispersion of $0.02$. 
In subsequent cell division cycles, the daughter cell properties are sampled from the same distribution as described above. 
At each time step, the growth rate of the cell is also chosen from a Gaussian distribution.  

{\bf Cell division:} The volume of growing cells increases at a constant rate $r_V$. 
Cell radii are updated from a Gaussian distribution with the mean rate $\dot{R} = (4\pi R_m^2)^{-1} r_V$, and dispersion of $10^{-7}$.  
Over the cell cycle time $\tau$, 
\begin{equation}
r_V = \frac{2\pi (R_{m})^3}{3\tau},
\end{equation}
where $R_{m}$ is the mitotic radius. Table I  lists  the parameters used in the simulations.


\newpage
\begin{table}\centering
\caption{Parameters used in the simulations. }

\begin{tabular}{ |p{7cm}||p{4cm}|p{5cm}|p{3cm}|  }
\hline
 \bf{Parameters} & \bf{Values} & \bf{References} \\
 \hline
Critical Radius for Division ($R_{m}$) &  5 $\mathrm{\mu m}$ & ~\cite{schaller2005multicellular}\\
 \hline
 Extracellular Matrix (ECM) Viscosity ($\eta$) & 0.005 $\mathrm{kg/ (\mu m~s)}$   & ~\cite{galle2005modeling}  \\
 \hline
 Benchmark Cell Cycle Time ($\tau_{min} = k_b^{-1}$)  & 900 $\mathrm{s}$  & ~\cite{kane1993zebrafish}\\
 \hline
 Adhesive Strength ($f^{ad})$&  $1\times 10^{-4} \mathrm{\mu N/\mu m^{2}}$  & ~\cite{schaller2005multicellular}, This paper \\
 \hline
Mean Cell Elastic Modulus ($E_{i}) $ & $10^{-3} \mathrm{MPa}$  & ~\cite{galle2005modeling}    \\
 \hline
Mean Cell Poisson Ratio ($\nu_{i}$) & 0.5 & ~\cite{schaller2005multicellular}  \\
 \hline
 Death Rate ($k_a$) & $10^{-6} \mathrm{s^{-1}}$ & This paper \\
 \hline
Mean Receptor Concentration ($c^{rec}$) & 0.90 (Normalized) & ~\cite{schaller2005multicellular} \\
\hline
Mean Ligand Concentration ($c^{lig}$) & 0.90 (Normalized) & ~\cite{schaller2005multicellular}  \\
\hline
Adhesive Friction $\gamma^{max}$ &  $10^{-4} \mathrm{kg/ (\mu m^{2}~s)}$  &  This paper\\
\hline
Threshold Pressue ($p_c$) & $1.0-3.5\times10^{-4} \mathrm{MPa}$  & ~\cite{schaller2005multicellular,montel2011stress}    \\
\hline
\end{tabular}
\end{table}

{\bf Cell-Cell Interactions:} 
By building on our previous work \cite{malmi2018cell}, we consider the cell-cell interactions to be driven by elastic (repulsive) and adhesive (attractive) forces.
 The total force on the $i^{th}$ cell is, 
$\vec{F}_{i} = \Sigma_{j \epsilon NN(i)}(F_{ij}^{el}-F_{ij}^{ad})\vec{n}_{ij}$, 
where $F_{ij}^{el}$ and $F_{ij}^{ad})$ are cell-cell elastic and 
adhesive forces respectively, and $\vec{n}_{ij}$ is the unit vector from the center of cell $j$ to cell $i$. 
We used the Hertz contact mechanics~\cite{schaller2005multicellular, drasdo2005single, pathmanathan2009computational} 
to model the elastic forces between the spherical cells.  $R_{i}$ and $R_{j}$, are the cell radii and 
the parameters $E_{i}$ and $\nu_{i}$, respectively, are the elastic 
modulus and 
Poisson ratio of the $i^{th}$ cell in the elastic force term:  $F_{ij}^{el}=\frac{h_{ij}^{3/2}}
{\frac{3}{4}(\frac{1-\nu_i^2}{E_i}+\frac{1-\nu_j^2}{E_j})\sqrt{\frac{1}{R_i(t)}+\frac{1}{R_j(t)}}}$. The overlap between two cells, $h_{ij}$, 
is defined as $\mathrm{max}[0, R_i + R_j - |\vec{r}_i - \vec{r}_j|]$. 

The inter-cell adhesive force, $F^{ad}_{ij}$, is given by,
\begin{equation}
\label{ad}
F_{ij}^{ad} = A_{ij}f^{ad}\frac{1}{2}(c_{i}^{rec}c_{j}^{lig} + c_{j}^{rec}c_{i}^{lig}). 
\end{equation}
In the above equation, $A_{ij}=\pi h_{ij} R_i R_j / (R_i + R_j)$ is the cell-cell contact area, $c_{i}^{rec}$ ($c_{i}^{lig}$) is the adhesion  receptor (ligand) concentration 
(assumed to be normalized with respect to the maximum receptor or ligand concentration such that  
$0 \leq c_{i}^{rec},  c_{i}^{lig} \leq 1$)~\cite{schaller2005multicellular}. The dimensions of the adhesive strength, $f^{ad}$,  in Eq.~\ref{ad} is $\mu N/ \mu m^2$. 

{\bf  Equations of Motion:}  If inertial effects are negligible~\cite{malmi2018cell}, the equation of motion of the $i^{th}$ cell is, 
\begin{equation}
\label{eqforce}
\dot{\vec{r}}_{i} = \frac{\vec{F}_{i}}{\gamma_i}, 
\end{equation}
where, $\gamma_i = \gamma_{i}^{\alpha' \beta', visc} + \gamma_{i}^{\alpha' \beta', ad} $ is the total friction coefficient. The cell-to-matrix friction is given by, 
\begin{equation}
\gamma_{i}^{\alpha' \beta', visc} = 6\pi \eta R_{i} \delta^{\alpha' \beta'}. 
\end{equation}
The cell-to-cell damping coefficient is,
 \begin{eqnarray}
\gamma_{i}^{\alpha' \beta', ad} =&& \gamma^{max}\Sigma_{j \epsilon NN(i)} (A_{ij}\frac{1}{2}(1+\frac{\vec{F}_{i} \cdot \vec{n}_{ij}}{|\vec{F}_{i}|})\times \\ \nonumber &&\frac{1}{2}(c_{i}^{rec}c_{j}^{lig} + c_{j}^{rec}c_{i}^{lig}))\delta^{\alpha' \beta'} \, .
\end{eqnarray}
The indices $\alpha'$ and $\beta'$ represent cartesian co-ordinates. 
Viscosity of the medium surrounding the cell is denoted by $\eta$ and $\gamma^{max}$ is the adhesive 
friction coefficient. 

{\bf Cell birth, apoptosis and dormancy :} We implement a mechanical feedback on cell growth and division on the basis of the local forces experienced by a cell. 
Depending on the local forces, a cell can either be in the 
dormant ($D$) or in the growth ($G$) phase. 
The effect of the local cellular microenvironment on proliferation, a collective cell mechanical effect, is taken into account through the pressure experienced by the cell ($p_i$). 
We refer to $p_{i}$ as pressure because it has the same dimensions, and models the mechanical sensitivity of cell proliferation to the local environment. The total pressure ($p_{i}$) on the $i^{th}$ cell is,
\begin{equation}
\label{pre}
p_i = \sum_{j \in NN(i)} \frac{|{F_{ij}}|}{A_{ij}}, 
\end{equation} 
The non-zero value of $p_c$ is the source of biomechanical feedback that regulates cell growth. 

There are several ways to compute pressure in cell collectives (for a review see Ref.~\cite{van2015simulating}). Here, we  define pressure following previous studies~\cite{schaller2005multicellular, byrne2009individual}. These studies produced quantitative agreement with experiments on the dynamics  
of multicellular spheroid growth~\cite{van2015simulating}.


\newpage
\bibliographystyle{naturemag}
\bibliography{supplement}